\documentclass[12pt,english,draftcls, one column]{IEEEtran}
\usepackage[T1]{fontenc}
\usepackage[latin9]{inputenc}
\usepackage{float}
\usepackage{amsmath}
\usepackage{amsthm}
\usepackage{amssymb}
\usepackage{graphicx}

\makeatletter

\floatstyle{ruled}
\newfloat{algorithm}{tbp}{loa}
\providecommand{\algorithmname}{Algorithm}
\floatname{algorithm}{\protect\algorithmname}

\theoremstyle{plain}

\theoremstyle{plain}

\ifCLASSINFOpdf% \usepackage[pdftex]{graphicx}
\else% or other class option (dvipsone, dvipdf, if not using dvips). graphicx
\fi% graphicx was written by David Carlisle and Sebastian Rahtz. It is
\usepackage{babel}

\makeatother

\usepackage{babel}
\providecommand{\propositionname}{Proposition}
\providecommand{\theoremname}{Theorem}

\begin{document}

\title{Joint Optimization of Cloud and Edge Processing for Fog Radio Access
Networks}

\author{Seok-Hwan Park, \textit{Member}, \textit{IEEE}, Osvaldo Simeone,
\textit{Fellow, IEEE}, \\ and Shlomo Shamai (Shitz),\textit{ Fellow,
IEEE} \thanks{S.-H. Park was supported by the National Research Foundation of Korea
(NRF) grant funded by the Korea government(Ministry of Science, ICT\&Future
Planning) {[}2015R1C1A1A01051825{]}. The work of O. Simeone was partially
supported by the U.S. NSF through grant 1525629. The work of S. Shamai
was partly supported by the Israel Science Foundation (ISF).

S.-H. Park is with the Division of Electronic Engineering, Chonbuk
National University, Jeonju 561-756, Korea (email: seokhwan@jbnu.ac.kr).

O. Simeone is with the Center for Wireless Communication and Signal
Processing Research, New Jersey Institute of Technology, 07102 Newark,
New Jersey, USA (email: osvaldo.simeone@njit.edu).

S. Shamai (Shitz) is with the Department of Electrical Engineering,
Technion, Haifa, 32000, Israel (email: sshlomo@ee.technion.ac.il).}}
\maketitle
\begin{abstract}
This work studies the joint design of cloud and edge processing for
the downlink of a fog radio access network (F-RAN). In an F-RAN, as
in cloud-RAN (C-RAN), a baseband processing unit (BBU) can perform
joint baseband processing on behalf of the remote radio heads (RRHs)
that are connected to the BBU by means of the fronthaul links. In
addition to the minimal functionalities of conventional RRHs in C-RAN,
the RRHs in an F-RAN may be equipped with local caches, in which frequently
requested contents can be stored, as well as with baseband processing
capabilities. They are hence referred to as enhanced RRH (eRRH). This
work focuses on the design of the delivery phase for an arbitrary
pre-fetching strategy used to populate the caches of the eRRHs. Two
fronthauling modes are considered, namely a \textit{hard-transfer
mode}, whereby non-cached files are communicated over the fronthaul
links to a subset of eRRHs, and a \textit{soft-transfer mode}, whereby
the fronthaul links are used to convey quantized baseband signals
as in a C-RAN. Unlike the hard-transfer mode in which baseband processing
is traditionally carried out only at the eRRHs, the soft-transfer
mode enables both centralized precoding at the BBU and local precoding
at the eRRHs based on the cached contents, by means of a novel superposition
coding approach. To attain the advantages of both approaches, a hybrid
design of soft- and hard-transfer modes is also proposed. The problem
of maximizing the delivery rate is tackled under fronthaul capacity
and per-eRRH power constraints. Numerical results are provided to
compare the performance of hard- and soft-transfer fronthauling modes,
as well as of the hybrid scheme, for different baseline pre-fetching
strategies. \end{abstract}


\theoremstyle{theorem}
\newtheorem{theorem}{Theorem}
\theoremstyle{proposition}
\newtheorem{proposition}{Proposition}
\theoremstyle{lemma}
\newtheorem{lemma}{Lemma}
\theoremstyle{corollary}
\newtheorem{corollary}{Corollary}
\theoremstyle{definition}
\newtheorem{definition}{Definition}
\theoremstyle{remark}
\newtheorem{remark}{Remark}

\section{Introduction}

Cloud radio access network (C-RAN) is an emerging architecture for
the fifth-generation (5G) of wireless system, in which a centralized
baseband signal processing unit (BBU) implements the baseband processing
functionalities of a set of remote radio heads (RRHs), which are connected
to the BBU by means of fronthaul links \cite{Checko}-\cite{Simeone-et-al}.
In the digital fronthauling adopted by the Common Public Radio Interface
(CPRI) specification \cite{CPRI}, the BBU quantizes and compresses
the encoded baseband signals prior to the transfer to the RRHs (see,
e.g., \cite{Simeone-et-al:ETT}-\cite{Patil-Yu}).

\begin{figure}
\centering\includegraphics[width=13.2cm,height=7.5cm]{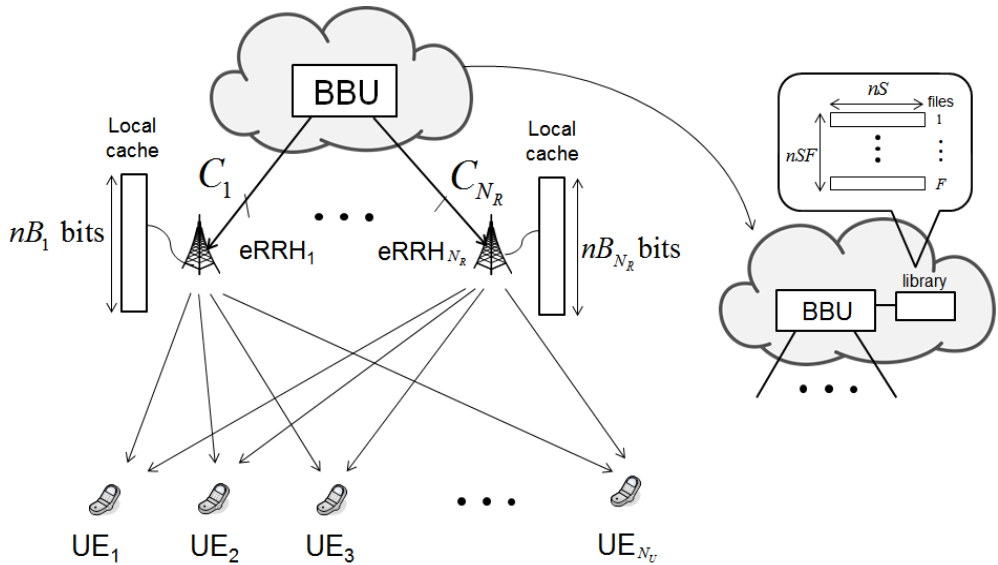}

\caption{\label{fig:System-Model}Illustration of an F-RAN, which has both
cloud and edge processing capabilities: the BBU, in the ``cloud'',
can perform joint baseband processing and the eRRHs are equipped with
local caches.}
\end{figure}

Recently, an evolved network architecture, referred to as \textit{Fog
Radio Access Network} (F-RAN), has been proposed, which enhances the
C-RAN architecture by allowing the RRHs to be equipped with storage
and signal processing functionalities \cite{Peng-et-al:FRAN}-\cite{China}.
The resulting RRHs are referred to here as \textit{enhanced RRHs}
(eRRHs)\footnote{In \cite{China}, eRRHs are referred to as Radio Remote Systems (RRSs).}.
In an F-RAN, edge caching can be performed to pre-fetch the most frequently
requested files to the eRRHs' local caches, as illustrated in Fig.
\ref{fig:System-Model}. In this way, fronthaul overhead can be reduced
and higher spectral efficiencies or lower delivery latency can be
obtained. It is emphasized that, unlike C-RAN \cite{China-CRAN},
the goal of the F-RAN architecture is not that of minimizing the deployment
and operating costs by means of reduced-complexity edge nodes, but
rather that of maximizing the system performance in terms of delivery
rate by leveraging both \textit{cloud (BBU) and edge (caching) resources}
\cite{Peng-et-al:cache-placement}-\cite{Ugur-et-al:ITA}\footnote{See also \cite[Sec. D]{Chiang}.}.

As a cache-aided system, an F-RAN operates in two phases, namely the
pre-fetching and the delivery phases \cite{Peng-et-al:cache-placement}-\cite{Ugur-et-al:ITA}
(see also \cite{MAli-Niesen}\cite{Sengupta-et-al}). Pre-fetching
operates at the large time scale corresponding to the period in which
content popularity remains constant. This time scale encompasses multiple
transmission intervals, as seen in Fig. \ref{fig:time-scale}. Based
on the cached file messages, the delivery phase, instead, operates
separately on each transmission interval.

\textbf{Related Works:} In \cite{Peng-et-al:cache-placement}, the
fronthaul-aware design of the pre-fetching policy was studied with
the aim of minimizing the average delivery latency while satisfying
the cache memory constraints. Since the optimization problem turns
out to be a mixed integer nonlinear program, the authors obtained
a difference-of-convex (DC) problem by means of smooth approximation
and integer relaxation, and proposed a successive convex approximation
algorithm. In \cite{Tao-et-al}, the authors consider the joint design
of cooperative beamforming and eRRH clustering for the delivery phase,
under an arbitrary fixed pre-fetching strategy, with the goal of minimizing
the network cost, which is defined as the sum of transmit power and
backhaul cost, under quality-of-service constraints. A similar problem
was tackled in \cite{Ugur-et-al} by assuming that coded, instead
of uncoded, caching is exploited (see also \cite{Bioglio-et-al}).
In \cite{Chen-et-al}, a stochastic geometry-based analysis is provided
of a specific hybrid caching strategy (see Sec. \cite[Sec. II-B]{Chen-et-al}).
Reference \cite{Azari-et-al} proposes a hypergraph-based framework
to obtain first-order quantitative insights into the performance of
an F-RAN architecture without the need to perform the non-convex optimization
studied in \cite{Peng-et-al:cache-placement}-\cite{Ugur-et-al}. An
information-theoretic framework for the analysis of latency in F-RANs
is developed in \cite{Tandon-Simeone}.

\textbf{Main Contributions:} In all the references \cite{Peng-et-al:cache-placement}-\cite{Tandon-Simeone}
summarized above, the fronthaul links in an F-RAN are leveraged in
a \textit{hard-transfer mode} to convey to the eRRHs the requested
content that is not present in the local caches. In contrast, in this
work, we consider not only the mentioned hard-transfer mode, but also
a novel \textit{soft-transfer mode} for the use of the fronthaul links.
The proposed approach is based on fronthaul quantization and superposition
coding: each eRRH transmits the superposition of two signals, one
that is locally encoded based on the content of the cache and another
that is encoded at the BBU and quantized for transmission on the fronthaul
link. Specifically, we study the joint design of cloud and edge processing
for the delivery phase of an F-RAN for an arbitrary pre-fetching strategy
by considering hard-transfer and soft-transfer fronthauling strategies.
For both fronthauling modes, we tackle the problem of optimizing cloud
and edge processing, i.e., processing at the BBU and at the eRRHs,
with the goal of maximizing the delivery rate while satisfying fronthaul
capacity and per-eRRH power constraints. Furthermore, to reap the
advantages of the two fronthauling approaches, we also propose a hybrid
design of hard- and soft-transfer modes, which is akin to \cite{Patil-Yu},
where it was studied in the absence of caching. Numerical results
are provided to compare the performance of hard- and soft-transfer
fronthauling modes, as well as the hybrid scheme, for baseline pre-fetching
strategies.

The rest of the paper is organized as follows. We describe the system
model in Sec. \ref{sec:System-Model} and review some baseline pre-fetching
strategies in Sec. \ref{sec:Pre-Fetching}. We discuss the design
of delivery phase under hard-transfer fronthaul mode in Sec. \ref{sec:Hard-Transfer}
and then propose a novel soft-transfer strategy in Sec. \ref{sec:Soft-Transfer}.
A hybrid design of hard- and soft-transfer modes is studied in Sec.
\ref{sec:Hybrid-Fronthauling}, and extensive numerical results are
presented in Sec. \ref{sec:Numerical-Results}. We close the paper
with some concluding remarks in Sec. \ref{sec:Conclustion}.

\textit{Notation}: We adopt standard information-theoretic definitions
for the mutual information $I(X;Y)$ between the random variables
$X$ and $Y$ \cite{ElGamal-Kim}. The circularly symmetric complex
Gaussian distribution with mean $\mbox{\boldmath${\mu}$}$ and covariance
matrix $\mathbf{R}$ is denoted by $\mathcal{CN}(\mbox{\boldmath${\mu}$},\bold{R})$.
The set of all $M\times N$ complex matrices is denoted by $\mathbb{C}^{M\times N}$,
and $\mathbb{E}(\cdot)$ represents the expectation operator. The
operation $(\cdot)^{\dagger}$ denotes Hermitian transpose of a matrix
or vector, and $\bar{a}$ is defined as $1-a$ for a binary variable
$a\in\{0,1\}$. For a scalar $x$, $\left\lfloor x\right\rfloor $
denotes the largest integer not larger than $x$.

\section{System Model\label{sec:System-Model}}

\begin{figure}
\centering\includegraphics[width=12cm,height=3cm]{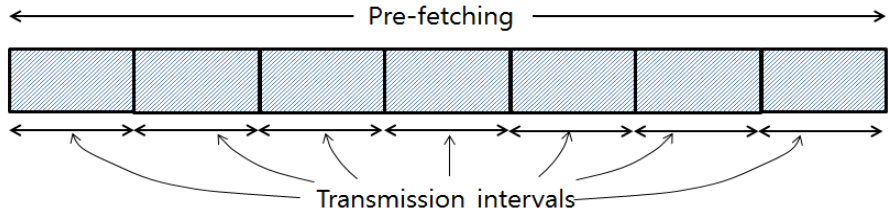}

\caption{\label{fig:time-scale}Illustration of the time scales of pre-fetching
and delivery phases.}
\end{figure}

As illustrated in Fig. \ref{fig:System-Model}, we consider the downlink
of an F-RAN, where $N_{U}$ multi-antenna user equipments (UEs) are
served by $N_{R}$ multi-antenna eRRHs that are connected to a BBU
in the ``cloud'' through digital fronthaul links. In addition to
the functionalities performed by conventional RRHs in C-RAN, such
as upconversion and RF transmission, each eRRH $i$ in an F-RAN is
equipped with a cache, which can store $nB_{i}$ bits, where $n$
is the number of (baud-rate) symbols of each downlink coded transmission
block. Furthermore, it also has baseband processing capabilities.
Each eRRH $i$ is connected to the BBU with a fronthaul link of capacity
$C_{i}$ bit per symbol of the downlink channel for $i\in\mathcal{N}_{R}\triangleq\{1,\ldots,N_{R}\}$.
We denote the numbers of antennas of eRRH $i$ and UE $k$ by $n_{R,i}$
and $n_{U,k}$, respectively, and define the notations $n_{R}\triangleq\sum_{i\in\mathcal{N}_{R}}n_{R,i}$
and $n_{U}\triangleq\sum_{k\in\mathcal{N}_{U}}n_{U,k}$.

We consider communication for content delivery via the outlined F-RAN
system. Accordingly, UEs request contents, or files, from a library
of $F$ files, each of size $nS$ bits, which are delivered by the
network across a number of transmission intervals (see Fig. \ref{fig:time-scale}).
Labeling the files in order of popularity, the probability $P(f)$
of a file $f$ to be selected is defined by Zipf's distribution (see,
e.g., \cite{Peng-et-al:cache-placement}-\cite{Ugur-et-al})
\begin{equation}
P(f)=cf^{-\gamma}\label{eq:Zipf-distribution}
\end{equation}
for $f\in\mathcal{F}\triangleq\{1,\ldots,F\}$, where $\gamma\geq0$
is a given popularity exponent and $c\geq0$ is set such that $\sum_{f\in\mathcal{F}}P(f)=1$.
Note that, as the exponent $\gamma$ increases, the popularity distribution
becomes more skewed towards the most popular files. Each UE $k$ requests
file $f_{k}\in\mathcal{F}$ with the probability (\ref{eq:Zipf-distribution}),
and the requested files $f_{k}$ are independent across the index
$k$.

Assuming flat-fading channel, the baseband signal $\mathbf{y}_{k}\in\mathbb{C}^{n_{U,k}\times1}$
received by UE $k$ in each transmission interval is given as
\begin{equation}
\mathbf{y}_{k}=\sum_{i\in\mathcal{N}_{R}}\mathbf{H}_{k,i}\mathbf{x}_{i}+\mathbf{z}_{k}=\mathbf{H}_{k}\mathbf{x}+\mathbf{z}_{k},\label{eq:received-signal}
\end{equation}
where $\mathbf{x}_{i}\in\mathbb{C}^{n_{R,i}\times1}$ is the baseband
signal transmitted by eRRH $i$ in a given downlink discrete channel
use, or symbol; $\mathbf{H}_{k,i}\in\mathbb{C}^{n_{U,k}\times n_{R,i}}$
denotes the channel response matrix from eRRH $i$ to UE $k$; $\mathbf{z}_{k}\in\mathbb{C}^{n_{U,k}\times1}$
is the additive noise distributed as $\mathbf{z}_{k}\sim\mathcal{CN}(\mathbf{0},\mathbf{\Sigma}_{\mathbf{z}_{k}})$
for some covariance matrix $\mathbf{\Sigma}_{\mathbf{z}_{k}}$; $\mathbf{H}_{k}\triangleq[\mathbf{H}_{k,1}\ldots\mathbf{H}_{k,N_{R}}]\in\mathbb{C}^{n_{U,k}\times n_{R}}$
collects the channel matrices $\mathbf{H}_{k,i}$ from each eRRH $i$
to any UE $k$; and $\mathbf{x}\triangleq[\mathbf{x}_{1};\ldots;\mathbf{x}_{N_{R}}]\in\mathbb{C}^{n_{R}\times1}$
is the signal transmitted by all the eRRHs. We assume that each eRRH
$i$ is subject to the average transmit power constraint stated as
\begin{equation}
\mathbb{E}\left\Vert \mathbf{x}_{i}\right\Vert ^{2}\leq P_{i}.\label{eq:per-eRRH-power-constraint}
\end{equation}
Furthermore, the channel matrices $\{\mathbf{H}_{k,i}\}_{k\in\mathcal{N}_{U},i\in\mathcal{N}_{R}}$
are assumed to remain constant during each transmission interval and
to be known to the BBU and eRRHs. The robust design with imperfect
CSI or via alternating distributed optimization \cite{Fouladgar-et-al}
is out of the scope of this work.

The system operates in two phases, namely pre-fetching and delivery
(see, e.g., \cite{MAli-Niesen}). Pre-fetching operates at a large
time scale corresponding to the period in which file popularity remains
constant. This time scale encompasses multiple transmission intervals
as illustrated in Fig. \ref{fig:time-scale}. The delivery phase operates
separately on each transmission interval. We assume that files are
transmitted in successive transmission intervals, until all current
requests are satisfied, i.e., UE $k$ successfully decodes the requested
file $f_{k}$ for all $k\in\mathcal{N}_{U}$. Then, new requests $\{f_{k}\}_{k\in\mathcal{N}_{U}}$
are considered and the corresponding files are transmitted.

In the \textbf{pre-fetching phase}, each eRRH $i$ downloads and stores
up to $nB_{i}$ bits from the library of files, which is of size $nSF$
bits (see Fig. \ref{fig:System-Model}). We define the \textit{fractional
caching capacity} $\mu_{i}$ of eRRH $i$ as
\begin{equation}
\mu_{i}\triangleq\frac{B_{i}}{SF}.\label{eq:fraction-cached}
\end{equation}
Accordingly, each eRRH can potentially store a fraction $\mu_{i}$
of each file (see \cite{Tandon-Simeone}\cite{MAli-Niesen}\cite{Sengupta-et-al}).
Different standard pre-fetching policies will be considered as detailed
in Sec. \ref{sec:Pre-Fetching}. Note that pre-fetching strategies
cannot be adapted to the channel matrices or requested file profile
$\{f_{k}\}_{k\in\mathcal{N}_{U}}$ in each transmission interval.

In the \textbf{delivery phase}, the eRRHs transmit in the downlink
in order to deliver the requested files $\mathcal{F}_{\mathrm{req}}\triangleq\cup_{k\in\mathcal{N}_{U}}\{f_{k}\}$
to the UEs. The transmitted signal $\mathbf{x}_{i}$ of each eRRH
$i$ is obtained as a function of the information stored in its local
cache, as well as of the information received from the BBU on the
fronthaul link. We consider two different approaches depending on
the type of the information transferred on the fronthaul links: \textit{hard-transfer
fronthauling} and \textit{soft-transfer fronthauling}. In the former,
the fronthaul links are used for the transfer of hard information
regarding the missing files that are not cached by the eRRHs as in
\cite{Peng-et-al:cache-placement}-\cite{Ugur-et-al}; while, with
the soft-transfer mode, the fronthaul links transfer quantized version
of the precoded signals for the missing files, in line with the C-RAN
paradigm. Soft- and hard-mode fronthauling strategies were compared
for C-RAN systems, i.e., with no caching, in terms of achievable rates
under an ergodic fading channel model in \cite{Kang-et-al} and in
terms of energy expenditure in \cite{Dai-Yu}. In the next sections,
we detail separately the pre-fetching and delivery phases. Moreover,
for the delivery phase, we will consider separately operations with
hard- and soft-transfer fronthauling, and also with a hybrid scheme
that combines the advantages of the two fronthauling approaches.

\section{Pre-Fetching Phase\label{sec:Pre-Fetching}}

The pre-fetching policy chooses $nB_{i}$ bits out of the library
of $nSF$ bits to be stored in the cache of eRRH $i$. Different policies
for caching can be considered, including coded caching \cite{Ugur-et-al}\cite{Bioglio-et-al}.
The pre-fetching strategy is determined based only on long-term state
information about the popularity distribution $P(f)$, as well as
on the cache memory sizes $\{B_{i}\}_{i\in\mathcal{N}_{R}}$, file
size $nS$ and the fronthaul capacities $\{C_{i}\}_{i\in\mathcal{N}_{R}}$.

\begin{figure}
\centering\includegraphics[width=12cm,height=7cm]{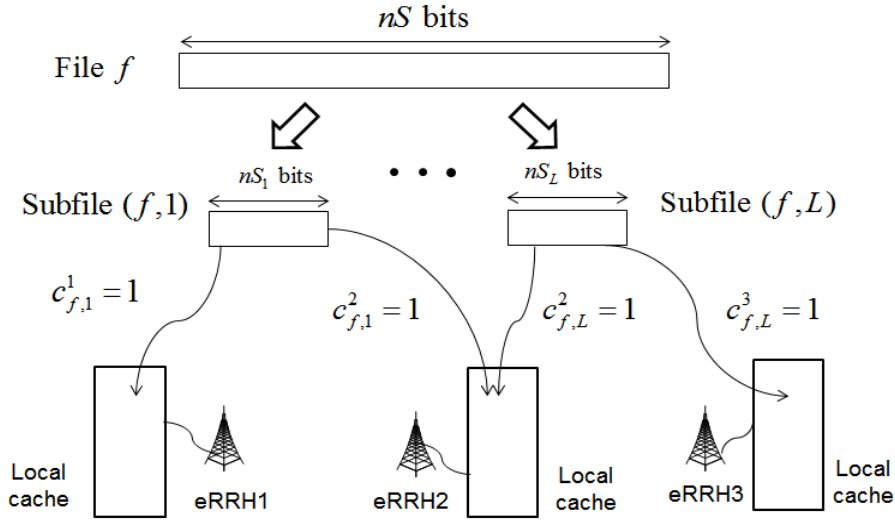}

\caption{\label{fig:Caching-variable}Illustration of the pre-fetching phase
for an example with $N_{R}=3$ eRRHs.}
\end{figure}

In this paper, as in \cite{Tao-et-al}\cite{Azari-et-al}\cite{MAli-Niesen},
we limit our attention to uncoded strategies. To this end, for the
sake of generality, we assume that each file $f$ is split into $L$
subfiles $(f,1),\ldots,(f,L)$ such that each subfile $(f,l)$ is
of size $nS_{l}$ bits with $\sum_{l\in\mathcal{L}}S_{l}=S$ and $\mathcal{L}\triangleq\{1,\ldots,L\}$
(see, e.g., \cite[Sec. III]{MAli-Niesen}). Then, the pre-fetching
strategy can be modeled by defining binary caching variables $\{c_{f,l}^{i}\}_{f\in\mathcal{F},l\in\mathcal{L},i\in\mathcal{N}_{R}}$
as
\begin{equation}
c_{f,l}^{i}=\begin{cases}
1, & \mathrm{if}\,\,\mathrm{subfile}\,\,(f,l)\,\,\mathrm{is}\,\,\mathrm{cached\,\,by\,\,eRRH}\,\,i\\
0, & \mathrm{otherwise}
\end{cases},\label{eq:caching-variable-definition}
\end{equation}
while satisfying the cache memory constraint at eRRH $i$ as
\begin{equation}
\sum_{f\in\mathcal{F}}\sum_{l\in\mathcal{L}}c_{f,l}^{i}S_{l}\leq B_{i}=\mu_{i}FS,\,\,\mathrm{for\,\,all}\,\,i\in\mathcal{N}_{R}.\label{eq:cache-memory-constraint}
\end{equation}
Fig. \ref{fig:Caching-variable} illustrates an example.

While the problem formulation to be given in later sections applies
to any choice of pre-fetching variables (\ref{eq:caching-variable-definition}),
the following subsections discuss three explicit standard pre-fetching
strategies that will be considered in Sec. \ref{sec:Numerical-Results}
for numerical performance evaluation. For the rest of this section,
we set $\mu_{i}=\mu$ for $i\in\mathcal{N}_{R}$ in order to avoid
a more cumbersome notation.

\subsection{Cache Most Popular\label{sub:Cache-Most-Popular}}

We first consider a pre-fetching strategy in which all eRRHs cache
the same $N_{C}$ most popular files, namely $f=1,\ldots,N_{C}$,
where $N_{C}$ is given as $N_{C}=\left\lfloor \mu F\right\rfloor $
in order to satisfy the cache constraints. This approach, which was
also considered in \cite[Sec. V]{Tao-et-al}, is expected to be a
good choice when the parameter $\gamma$ of the distribution $P(f)$
is large, i.e., when only a few popular files are frequently requested
by UEs. We obtain it by setting $L=1$ and
\begin{equation}
c_{f,l}^{i}=\begin{cases}
1, & \mathrm{if}\,\,f\leq N_{C}\\
0, & \mathrm{otherwise}
\end{cases}.\label{eq:caching-variable-CMP}
\end{equation}
We refer to this strategy as Cache Most Popular (CMP).

\subsection{Cache Distinct\label{sub:Cache-Distinct}}

When the parameter $\gamma$ is small, it may be advantageous to store
as many distinct files as possible in the caches. Thus, we also consider
a pre-fetching strategy where eRRH 1 stores files $1,N_{R}+1,\ldots$;
eRRH 2 stores files $2,N_{R}+2,\ldots$; and so on, until caches are
full. This pre-fetching strategy, referred to as Cache Distinct (CD),
is obtained by choosing $L=1$ and
\begin{equation}
c_{f,l}^{i}=\begin{cases}
1, & \mathrm{if}\,\,i=\mathrm{mod}(f-1,N_{R})+1\\
0, & \mathrm{otherwise}
\end{cases}.\label{eq:caching-variable-CD}
\end{equation}
The number $N_{C}$ of files that can be stored in each cache is again
$N_{C}=\left\lfloor \mu F\right\rfloor $.

\subsection{Fractional Cache Distinct\label{sub:Fractional-Cache-Distinct}}

Unlike CMP, CD does not enable cooperative transmission from multiple
eRRHs based only on the content of the caches, since each file cannot
be stored by multiple eRRHs. To address this issue, which can be significant
if the fronthaul capacities $C_{i}$ are small, we consider a Fractional
Cache Distinct (FCD) pre-fetching strategy, where each file $f$ is
split into multiple subfiles, i.e., $L>1$, and distributed over the
eRRHs as described below.

\subsubsection{Partial Caching ($\mu\leq1/N_{R}$)}

In this case, there is not enough caching capacity to store all files.
Each file $f$ is then split into $N_{R}+1$ disjoint subfiles, i.e.,
$L=N_{R}+1$, so that the first $N_{R}$ fragments $(f,1),\ldots,(f,N_{R})$
are distributed over eRRHs chosen randomly without replacement, while
the last fragment $(f,N_{R}+1)$ is not cached. To this end, the sizes
of the files are set to $S_{l}=\mu S$ and $S_{l}=(1-N_{R}\mu)S$
for $l\in\mathcal{N}_{R}$ and $l=N_{R}+1$, respectively. This policy
can be implemented by setting the caching variables $c_{f,l}^{i}$
to
\begin{equation}
c_{f,l}^{i}=\begin{cases}
1, & \mathrm{if}\,\,l=i_{f,l}\\
0, & \mathrm{otherwise}
\end{cases},\label{eq:caching-variable-FCD}
\end{equation}
where $i_{f,1},\ldots,i_{f,L}$ are obtained as random permutations
of the numbers $1,\ldots,N_{R}$, which are independent across the
file index $f$. Randomized caching was also considered in \cite[Sec. V]{Tao-et-al}
without file splitting, i.e., with $L=1$.

\subsubsection{Redundant Caching ($\mu>1/N_{R}$)}

In this case, eRRHs can potentially store overlapping fragments of
all files. Each file $f$ is split into $N_{R}$ disjoint subfiles,
i.e., $L=N_{R}$, each of equal size $S_{l}=S/N_{R}$. Each cache
can hence store up to $\left\lfloor \mu N_{R}\right\rfloor $ segments
of each file. To populate the caches, we divide each cache into $\left\lfloor \mu N_{R}\right\rfloor $
disjoint parts each of $nS_{l}$ bits. Each part $j=1,\ldots,\left\lfloor \mu N_{R}\right\rfloor $
across all eRRHs is populated by means of a random permutation of
the eRRHs' indices for each file as discussed above, with the caveat
that we exclude permutations by which an eRRH would store a segment
$(f,l)$ more than once.

We close this section with two remarks. First, a hybrid CMP and FCD
caching policy was proposed in \cite{Chen-et-al}, whereby part of
the cache of each eRRH is used to cache the same most popular files
and the rest is instead leveraged to store distinct fragments of less
popular files. The second remark is that the optimization of pre-fetching
strategy based on long-term state information could be addressed by
adopting stochastic optimization techniques (see, e.g., \cite{Razaviyayn-et-al}),
but here we leave this challenging aspect as an interesting open problem.

\section{Delivery Phase With Hard-Transfer Fronthauling\label{sec:Hard-Transfer}}

For a given pre-fetching strategy, in this section, we consider the
design of the delivery phase in each transmission interval under the
hard-transfer fronthaul mode, where the fronthaul links are used to
transfer hard information of subfiles that are not cached by eRRHs.
This mode was also considered in \cite{Peng-et-al:cache-placement}-\cite{Ugur-et-al}.
The formulation considered here is akin to that of \cite{Tao-et-al},
with the difference that in this paper we study the maximization of
the delivery rate under fronthaul capacity constraints, rather than
the minimization of a compound cost function that includes both downlink
power and fronthaul capacity as in \cite{Tao-et-al}. The analysis
of hard-mode fronthaul is included here mostly for the purpose of
comparison with the soft-transfer mode.

We allow any subfile $(f,l)$ to be delivered to the UE at a rate
$R_{f,l}\leq S_{l}$, so that $nR_{f,l}\leq nS_{l}$ bits are transmitted
to the UE in the given transmission interval. The remaining $nS_{l}-nR_{f,l}$
bits can then be sent in the following transmission intervals by solving
a similar optimization problem. Our goal is that of maximizing the
rates $R_{f,l}$ that can be transmitted on a per-transmission interval
basis.

Hard-mode fronthauling requires the determination of the set of eRRHs
to which each subfile $(f,l)$ is transferred on the fronthaul link.
We do this by defining the binary variable $d_{f,l}^{i}$ as
\begin{equation}
d_{f,l}^{i}=\begin{cases}
1, & \mathrm{if}\,\,\mathrm{subfile}\,\,(f,l)\,\,\mathrm{is}\,\,\mathrm{transferred\,\,to\,\,eRRH}\,\,i\\
0, & \mathrm{otherwise}
\end{cases}.\label{eq:assignment-variable-definition}
\end{equation}
The fronthaul capacity constraint for each eRRH $i$ is stated as
\begin{equation}
\sum_{f\in\mathcal{F}}\sum_{l\in\mathcal{L}}d_{f,l}^{i}R_{f,l}\leq C_{i}.\label{eq:fronthaul-capacity-constraint-hard-transfer}
\end{equation}

Based on the cached or transferred subfiles $(f,l)$ with $c_{f,l}^{i}=1$
or $d_{f,l}^{i}=1$, respectively, each eRRH $i$ performs channel
encoding to produce the encoded baseband signal $\mathbf{x}_{i}$.
Denoting as $\mathcal{N}_{F,i}\triangleq\{(f,l)|c_{f,l}^{i}=1\,\,\mathrm{or}\,\,d_{f,l}^{i}=1\}$
the set of subfiles available at eRRH $i$, the eRRH performs linear
precoding as in \cite{Tao-et-al} to obtain the transmitted signal
$\mathbf{x}_{i}$ as
\begin{equation}
\mathbf{x}_{i}=\sum_{(f,l)\in\mathcal{N}_{F,i}}\mathbf{V}_{f,l}^{i}\mathbf{s}_{f,l}=\sum_{f\in\mathcal{F}}\sum_{l\in\mathcal{L}}(1-\bar{c}_{f,l}^{i}\bar{d}_{f,l}^{i})\mathbf{V}_{f,l}^{i}\mathbf{s}_{f,l},\label{eq:linear-precoding-hard}
\end{equation}
where $\mathbf{V}_{f,l}^{i}\in\mathbb{C}^{n_{R,i}\times n_{S,f,l}}$
is the precoding matrix for the baseband signal $\mathbf{s}_{f,l}\in\mathbb{C}^{n_{S,f,l}\times1}$
that encodes the subfile $(f,l)$ and is distributed as $\mathbf{s}_{f,l}\sim\mathcal{CN}(\mathbf{0},\mathbf{I})$.

With (\ref{eq:linear-precoding-hard}), the received signal $\mathbf{y}_{k}$
in (\ref{eq:received-signal}) can be written as
\begin{equation}
\mathbf{y}_{k}=\sum_{l\in\mathcal{L}}\mathbf{H}_{k}\bar{\mathbf{V}}_{f_{k},l}\mathbf{s}_{f_{k},l}+\sum_{f\in\mathcal{F}_{\mathrm{req}}\setminus\{f_{k}\}}\sum_{l\in\mathcal{L}}\mathbf{H}_{k}\bar{\mathbf{V}}_{f,l}\mathbf{s}_{f,l}+\mathbf{z}_{k},\label{eq:received-signal-hard}
\end{equation}
where the aggregated precoding matrix $\bar{\mathbf{V}}_{f,l}\in\mathbb{C}^{n_{R}\times n_{S,f,l}}$
for subfile $(f,l)$ is defined as
\begin{equation}
\bar{\mathbf{V}}_{f,l}\triangleq\left[(1-\bar{c}_{f,l}^{1}\bar{d}_{f,l}^{1})\mathbf{V}_{f,l}^{1};\,(1-\bar{c}_{f,l}^{2}\bar{d}_{f,l}^{2})\mathbf{V}_{f,l}^{2};\,\ldots\,;\,(1-\bar{c}_{f,l}^{N_{R}}\bar{d}_{f,l}^{N_{R}})\mathbf{V}_{f,l}^{N_{R}}\right].\label{eq:aggregated-precoding-matrix-hard}
\end{equation}
In (\ref{eq:received-signal-hard}), the first term is the desired
signal to be decoded by the receiving UE $k$, and the second term
is the superposition of the interference signals encoding the files
requested by the other UEs.

We assume that, based on (\ref{eq:received-signal-hard}), each UE
$k$ performs successive interference cancellation (SIC) decoding.
Without loss of generality, we consider the decoding order $\mathbf{s}_{f_{k},1}\rightarrow\ldots\rightarrow\mathbf{s}_{f_{k},L}$
so that the rate $R_{f_{k},l}$ of the subfile $(f_{k},l)$ is bounded
as
\begin{align}
R_{f_{k},l} & \leq q_{k,l}\left(\bar{\mathbf{V}}\right)\label{eq:achievable-rate-hard}\\
 & \triangleq I\left(\mathbf{s}_{f_{k},l};\mathbf{y}_{k}|\mathbf{s}_{f_{k},1},\ldots,\mathbf{s}_{f_{k},l-1}\right)\nonumber \\
 & =\log\det\left(\sum_{m=l}^{L}\mathbf{H}_{k}\bar{\mathbf{V}}_{f_{k},m}\bar{\mathbf{V}}_{f_{k},m}^{\dagger}\mathbf{H}_{k}^{\dagger}+\sum_{f\in\mathcal{F}_{\mathrm{req}}\setminus\{f_{k}\}}\sum_{m\in\mathcal{L}}\mathbf{H}_{k}\bar{\mathbf{V}}_{f,m}\bar{\mathbf{V}}_{f,m}^{\dagger}\mathbf{H}_{k}^{\dagger}+\mathbf{\Sigma}_{\mathbf{z}_{k}}\right)\nonumber \\
 & -\log\det\left(\sum_{m=l+1}^{L}\mathbf{H}_{k}\bar{\mathbf{V}}_{f_{k},m}\bar{\mathbf{V}}_{f_{k},m}^{\dagger}\mathbf{H}_{k}^{\dagger}+\sum_{f\in\mathcal{F}_{\mathrm{req}}\setminus\{f_{k}\}}\sum_{m\in\mathcal{L}}\mathbf{H}_{k}\bar{\mathbf{V}}_{f,m}\bar{\mathbf{V}}_{f,m}^{\dagger}\mathbf{H}_{k}^{\dagger}+\mathbf{\Sigma}_{\mathbf{z}_{k}}\right),\nonumber
\end{align}
where we defined the notation $\bar{\mathbf{V}}\triangleq\{\bar{\mathbf{V}}_{f,l}\}_{f\in\mathcal{F}_{\mathrm{req}},\,l\in\mathcal{L}}$.

\subsection{Problem Definition and Optimization\label{sub:Problem-Formulation-Hard}}

We aim at maximizing the minimum-user rate $R_{\min}$ defined as
$R_{\min}\triangleq\min_{f\in\mathcal{F}_{\mathrm{req}}}R_{f}$ while
satisfying per-eRRH fronthaul capacity and power constraints, where
$R_{f}\triangleq\sum_{l\in\mathcal{L}}R_{f,l}$ denotes the achievable
delivery rate for file $f$. We recall from our discussion above that
maximizing $R_{\min}$ is instrumental in reducing the number of transmission
intervals needed to deliver all the files $\mathcal{F}_{\mathrm{req}}$
to the requesting UEs. The problem is stated as\begin{subequations}\label{eq:problem-hard-original}
\begin{align}
\underset{\bar{\mathbf{V}},R_{\min},\mathbf{R}}{\mathrm{maximize}}\,\, & R_{\min}\label{eq:problem-hard-original-objective}\\
\mathrm{s.t.}\,\,\, & R_{\min}\leq\sum_{l\in\mathcal{L}}R_{f,l},\,\,f\in\mathcal{F}_{\mathrm{req}},\label{eq:problem-hard-original-rmin-constraint}\\
 & R_{f_{k},l}\leq q_{k,l}\left(\bar{\mathbf{V}}\right),\,\,l\in\mathcal{L},\,k\in\mathcal{N}_{U},\label{eq:problem-hard-original-rate-constraint}\\
 & \sum_{f\in\mathcal{F}_{\mathrm{req}}}\sum_{l\in\mathcal{L}}d_{f,l}^{i}R_{f,l}^{i}\leq C_{i},\,\,i\in\mathcal{N}_{R},\label{eq:problem-hard-original-fronthaul-constraint}\\
 & R_{f,l}\leq S_{l},\,\,f\in\mathcal{F}_{\mathrm{req}},\,l\in\mathcal{L},\label{eq:problem-hard-original-file-size-constraint}\\
 & \sum_{f\in\mathcal{F}_{\mathrm{req}}}\sum_{l\in\mathcal{L}}(1-\bar{c}_{f,l}^{i}\bar{d}_{f,l}^{i})\mathrm{tr}\left(\mathbf{E}_{i}^{\dagger}\bar{\mathbf{V}}_{f,l}\bar{\mathbf{V}}_{f,l}^{\dagger}\mathbf{E}_{i}\right)\leq P_{i},\,\,i\in\mathcal{N}_{R},\label{eq:problem-hard-original-power-constraint}
\end{align}
\end{subequations}where we define the matrix $\mathbf{E}_{i}\in\mathbb{C}^{n_{R}\times n_{R,i}}$
containing zero entries except for the rows from $\sum_{j=1}^{i-1}n_{R,j}+1$
to $\sum_{j=1}^{i}n_{R,j}$ containing the identity matrix of size
$n_{R,i}$, and the notation $\mathbf{R}\triangleq\{R_{f,l}\}_{f\in\mathcal{F}_{\mathrm{req}},\,l\in\mathcal{L}}$.
In the problem, the constraint (\ref{eq:problem-hard-original-file-size-constraint})
imposes that the rate $R_{f,l}$ of each subfile be limited by the
subfile size $S_{l}$, and the constraint (\ref{eq:problem-hard-original-power-constraint})
is equivalent to the per-eRRH power constraints (\ref{eq:per-eRRH-power-constraint})
within the precoding model (\ref{eq:linear-precoding-hard}). We emphasize
that in (\ref{eq:problem-hard-original}), the pre-fetching variables
(\ref{eq:caching-variable-definition}) and the fronthaul transfer
variables (\ref{eq:assignment-variable-definition}) are fixed.

The solution of problem (\ref{eq:problem-hard-original}) is made
difficult by the non-convexity in the constraint (\ref{eq:problem-hard-original-rate-constraint}).
Here, noting that the left-hand side of (\ref{eq:problem-hard-original-rate-constraint})
has the DC structure when stated in terms of the covariance matrices
$\mathbf{W}_{f,l}\triangleq\bar{\mathbf{V}}_{f,l}\bar{\mathbf{V}}_{f,l}^{\dagger}\succeq\mathbf{0}$,
as in \cite{Park-et-al:TSP13}\cite{Park-et-al:SPM}\cite{Tao-et-al},
we adopt the concave-convex procedure (CCCP) for tackling (\ref{eq:problem-hard-original}).
Specifically, we address problem (\ref{eq:problem-hard-original})
with optimization variables $\mathbf{W}\triangleq\{\mathbf{W}_{f,l}\}_{f\in\mathcal{F}_{\mathrm{req}},\,l\in\mathcal{L}}$
by relaxing the rank constraints $\mathrm{rank}(\mathbf{W}_{f,l})\leq n_{S,f,l}$.

The resulting algorithm is described in Algorithm 1, where the function
$\tilde{q}_{k,l}(\mathbf{W}^{(t+1)},\mathbf{W}^{(t)})$ is defined
as
\begin{align}
 & \tilde{q}_{k,l}\left(\mathbf{W}^{(t+1)},\mathbf{W}^{(t)}\right)\label{eq:achievable-rate-linearized-hard}\\
\triangleq & \log\det\left(\sum_{m=l}^{L}\mathbf{H}_{k}\mathbf{W}_{f_{k},m}^{(t+1)}\mathbf{H}_{k}^{\dagger}+\sum_{f\in\mathcal{F}_{\mathrm{req}}\setminus\{f_{k}\}}\sum_{m\in\mathcal{L}}\mathbf{H}_{k}\mathbf{W}_{f,m}^{(t+1)}\mathbf{H}_{k}^{\dagger}+\mathbf{\Sigma}_{\mathbf{z}_{k}}\right)\nonumber \\
- & \varphi\left(\begin{array}{c}
\sum_{m=l+1}^{L}\mathbf{H}_{k}\mathbf{W}_{f_{k},m}^{(t+1)}\mathbf{H}_{k}^{\dagger}+\sum_{f\in\mathcal{F}_{\mathrm{req}}\setminus\{f_{k}\}}\sum_{m\in\mathcal{L}}\mathbf{H}_{k}\mathbf{W}_{f,m}^{(t+1)}\mathbf{H}_{k}^{\dagger}+\mathbf{\Sigma}_{\mathbf{z}_{k}},\\
\sum_{m=l+1}^{L}\mathbf{H}_{k}\mathbf{W}_{f_{k},m}^{(t)}\mathbf{H}_{k}^{\dagger}+\sum_{f\in\mathcal{F}_{\mathrm{req}}\setminus\{f_{k}\}}\sum_{m\in\mathcal{L}}\mathbf{H}_{k}\mathbf{W}_{f,m}^{(t)}\mathbf{H}_{k}^{\dagger}+\mathbf{\Sigma}_{\mathbf{z}_{k}}
\end{array}\right),\nonumber
\end{align}
with the notation $\varphi(\mathbf{A},\mathbf{B})\triangleq\log\det(\mathbf{B})+\mathrm{tr}(\mathbf{B}^{-1}(\mathbf{A}-\mathbf{B}))$.
After the convergence of the algorithm, each precoding matrix $\bar{\mathbf{V}}_{f,l}$
is obtained as $\bar{\mathbf{V}}_{f,l}\leftarrow\mathbf{V}_{n_{S,f,l}}(\mathbf{W}_{f,l})\mathrm{diag}(\mathbf{d}_{n_{S,f,l}}(\mathbf{W}_{f,l}))^{1/2}$,
where $\mathbf{V}_{N}(\mathbf{A})$ takes the $N$ leading eigenvectors
of the matrix $\mathbf{A}$ as its columns, $\mathbf{d}_{N}(\mathbf{A})$
is a vector whose elements are given as the corresponding eigenvalues,
and each precoding matrix $\mathbf{V}_{f,l}^{i}$ for eRRH $i$ can
be obtained as $\mathbf{V}_{f,l}^{i}\leftarrow(1-\bar{c}_{f,l}^{i}\bar{d}_{f,l}^{i})\mathbf{E}_{i}^{\dagger}\bar{\mathbf{V}}_{f,l}$.
We refer to \cite{Tao-et-al} for a discussion of known results on
the convergence of CCCP. We also note that, an alternative approach,
not based on rank relaxation, would be to use successive convex approximation
methods \cite{Scutari} based on lower bounds obtained from Fenchel
duality (see, e.g., \cite{Borwein-Lewis}).

\begin{algorithm}
\caption{CCCP algorithm for problem (\ref{eq:problem-hard-original})}

\textbf{1.} Initialize the matrices $\mathbf{W}^{(1)}$ to arbitrary
positive semidefinite matrices that satisfy the per-eRRH power constraints
(\ref{eq:problem-hard-original-power-constraint}) and set $t=1$.

\textbf{2.} Update the matrices $\mathbf{W}^{(t+1)}$ as a solution
of the following convex problem:\begin{subequations}\label{eq:problem-hard-DC}
\begin{align}
\underset{\mathbf{W}^{(t+1)}\succeq\mathbf{0},R_{\min},\mathbf{R}}{\mathrm{maximize}}\,\, & R_{\min}\label{eq:problem-hard-DC-objective}\\
\mathrm{s.t.}\,\,\, & R_{\min}\leq\sum_{l\in\mathcal{L}}R_{f,l},\,\,f\in\mathcal{F}_{\mathrm{req}},\label{eq:problem-hard-DC-rmin-constraint}\\
 & R_{f_{k},l}\leq\tilde{q}_{k,l}\left(\mathbf{W}^{(t+1)},\mathbf{W}^{(t)}\right),\,\,l\in\mathcal{L},\,k\in\mathcal{N}_{U},\label{eq:problem-hard-DC-rate-constraint}\\
 & \sum_{f\in\mathcal{F}_{\mathrm{req}}}\sum_{l\in\mathcal{L}}d_{f,l}^{i}R_{f,l}^{i}\leq C_{i},\,\,i\in\mathcal{N}_{R},\label{eq:problem-hard-DC-fronthaul-constraint}\\
 & R_{f,l}\leq S_{l},\,\,f\in\mathcal{F}_{\mathrm{req}},\,l\in\mathcal{L},\label{eq:problem-hard-DC-file-size-constraint}\\
 & \sum_{f\in\mathcal{F}_{\mathrm{req}}}\sum_{l\in\mathcal{L}}(1-\bar{c}_{f,l}^{i}\bar{d}_{f,l}^{i})\mathrm{tr}\left(\mathbf{E}_{i}^{\dagger}\mathbf{W}_{f,l}^{(t+1)}\mathbf{E}_{i}\right)\leq P_{i},\,\,i\in\mathcal{N}_{R}.\label{eq:problem-hard-DC-power-constraint}
\end{align}
\end{subequations}

\textbf{3.} Stop if a convergence criterion is satisfied. Otherwise,
set $t\leftarrow t+1$ and go back to Step 2.
\end{algorithm}

\section{Delivery Phase With Soft-Transfer Fronthauling\label{sec:Soft-Transfer}}

Unlike the hard-transfer mode that uses the fronthaul links to transfer
hard information on missing files, in the soft-transfer mode typical
of C-RAN, the fronthaul links are used to transfer a quantized version
of the precoded signals of the missing files. Accordingly, the signal
$\mathbf{x}_{i}$ transmitted by eRRH $i$ on the downlink channel
is given as the superposition of two signals, one that is locally
encoded based on the content in the cache and another that is encoded
at the BBU and quantized for transmission on the fronthaul link. This
yields
\begin{equation}
\mathbf{x}_{i}=\sum_{f\in\mathcal{F}_{\mathrm{req}}}\sum_{l\in\mathcal{L}}c_{f,l}^{i}\mathbf{V}_{f,l}^{i}\mathbf{s}_{f,l}+\hat{\mathbf{x}}_{i},\label{eq:transmitted-signal-soft-transfer}
\end{equation}
where $\mathbf{V}_{f,l}^{i}\in\mathbb{C}^{n_{R,i}\times n_{S,f,l}}$
is the precoding matrix for the baseband signal $\mathbf{s}_{f,l}$
encoding the cached file $(f,l)$, while $\hat{\mathbf{x}}_{i}$ represents
the quantized baseband signal received from the BBU on the fronthaul
link. Note that in a C-RAN, the transmitted signal would be given
solely by the quantized signal $\hat{\mathbf{x}}_{i}$, which is discussed
next.

The BBU precodes the subfiles that are not stored in each eRRH $i$
producing the signal
\begin{equation}
\tilde{\mathbf{x}}_{i}=\sum_{f\in\mathcal{F}_{\mathrm{req}}}\sum_{l\in\mathcal{L}}\bar{c}_{f,l}^{i}\mathbf{U}_{f,l}^{i}\mathbf{s}_{f,l},\label{eq:precoding-BBU-soft}
\end{equation}
where $\mathbf{U}_{f,l}^{i}\in\mathbb{C}^{n_{R,i}\times n_{S,f,l}}$
is the precoding matrix for the baseband signal $\mathbf{s}_{f,l}$
that encodes the fragment $(f,l)$ not available at eRRH $i$. The
signal (\ref{eq:precoding-BBU-soft}) is quantized, obtaining the
signal $\hat{\mathbf{x}}_{i}$ as
\begin{equation}
\hat{\mathbf{x}}_{i}=\tilde{\mathbf{x}}_{i}+\mathbf{q}_{i},\label{eq:quantized-signal-soft}
\end{equation}
where $\mathbf{q}_{i}$ denotes the quantization noise independent
of $\tilde{\mathbf{x}}_{i}$ and distributed as $\mathbf{q}_{i}\sim\mathcal{CN}(\mathbf{0},\mathbf{\Omega}_{i})$
with the covariance matrix $\mathbf{\Omega}_{i}\succeq\mathbf{0}$.
The signals $\tilde{\mathbf{x}}_{i}$ and $\tilde{\mathbf{x}}_{j}$
for different eRRHs $i\neq j$ are quantized independently so that
the quantization noise signals $\mathbf{q}_{i}$ and $\mathbf{q}_{j}$
are independent \cite{Simeone-et-al:ETT}\footnote{The multivariate compression method proposed in \cite{Park-et-al:TSP13}
allows the signals $\tilde{\mathbf{x}}_{i}$ and $\tilde{\mathbf{x}}_{j}$
for different eRRHs $i\neq j$ to be jointly quantized, hence obtaining
correlated quantization noises. We do not further pursue the application
of multivariate compression here, although its inclusion in the analysis
could be carried out in a similar manner.}. Using standard information theoretic results (see, e.g., \cite[Ch. 3]{ElGamal-Kim}),
the signal $\hat{\mathbf{x}}_{i}$ can be reliably recovered by eRRH
$i$ if the condition
\begin{align}
g_{i}\left(\mathbf{U},\mathbf{\Omega}\right) & \triangleq I\left(\tilde{\mathbf{x}}_{i};\hat{\mathbf{x}}_{i}\right)\label{eq:fronthaul-constraint-soft}\\
 & =\log\det\left(\sum_{f\in\mathcal{F}_{\mathrm{req}}}\sum_{l\in\mathcal{L}}\bar{c}_{f,l}^{i}\mathbf{U}_{f,l}^{i}\mathbf{U}_{f,l}^{i\dagger}+\mathbf{\Omega}_{i}\right)-\log\det\left(\mathbf{\Omega}_{i}\right)\leq C_{i}\nonumber
\end{align}
is satisfied, where we define the notations $\mathbf{U}\triangleq\{\mathbf{U}_{f,l}^{i}\}_{f\in\mathcal{F}_{req},l\in\mathcal{L},i\in\mathcal{N}_{R}}$
and $\mathbf{\Omega}\triangleq\{\mathbf{\Omega}_{i}\}_{i\in\mathcal{N}_{R}}$.

With (\ref{eq:transmitted-signal-soft-transfer}), the signal $\mathbf{y}_{k}$
received by UE $k$ in (\ref{eq:received-signal}) can be written
as
\begin{align}
\mathbf{y}_{k} & =\sum_{l\in\mathcal{L}}\mathbf{H}_{k}\bar{\mathbf{V}}_{f_{k},l}\mathbf{s}_{f_{k},l}+\sum_{f\in\mathcal{F}_{\mathrm{req}}\setminus\{f_{k}\}}\sum_{l\in\mathcal{L}}\mathbf{H}_{k}\bar{\mathbf{V}}_{f,l}\mathbf{s}_{f,l}+\mathbf{H}_{k}\mathbf{q}+\mathbf{z}_{k},\label{eq:received-signal-soft}
\end{align}
where we defined the aggregated precoding matrix $\bar{\mathbf{V}}_{f,l}\triangleq[\bar{\mathbf{V}}_{f,l}^{1};\ldots;\bar{\mathbf{V}}_{f,l}^{N_{R}}]$
for subfile $(f,l)$ with $\bar{\mathbf{V}}_{f,l}^{i}\triangleq c_{f,l}^{i}\mathbf{V}_{f,l}^{i}+\bar{c}_{f,l}^{i}\mathbf{U}_{f,l}^{i}$
and the quantization noise vector $\mathbf{q}\triangleq[\mathbf{q}_{1};\ldots;\mathbf{q}_{N_{R}}]$
distributed as $\mathbf{q}\sim\mathcal{CN}(\mathbf{0},\bar{\mathbf{\Omega}})$
with $\bar{\mathbf{\Omega}}\triangleq\mathrm{diag}(\mathbf{\Omega}_{1},\ldots,\mathbf{\Omega}_{N_{R}})$.
Similar to the case with hard-transfer fronthauling, we assume that
UE $k$ performs SIC decoding based on (\ref{eq:received-signal-soft})
with the decoding order $\mathbf{s}_{f_{k},1}\rightarrow\ldots\rightarrow\mathbf{s}_{f_{k},L}$,
so that the rate $R_{f_{k},l}$ of the subfile $(f_{k},l)$ is bounded
as
\begin{align}
R_{f_{k},l} & \leq q_{k,l}\left(\bar{\mathbf{V}},\mathbf{\Omega}\right)\label{eq:achievable-rate-soft}\\
 & \triangleq I\left(\mathbf{s}_{f_{k},l};\mathbf{y}_{k}|\mathbf{s}_{f_{k},1},\ldots,\mathbf{s}_{f_{k},l-1}\right)\nonumber \\
 & =\log\det\left(\sum_{m=l}^{L}\mathbf{H}_{k}\bar{\mathbf{V}}_{f_{k},m}\bar{\mathbf{V}}_{f_{k},m}^{\dagger}\mathbf{H}_{k}^{\dagger}+\sum_{f\in\mathcal{F}_{\mathrm{req}}\setminus\{f_{k}\}}\sum_{m\in\mathcal{L}}\mathbf{H}_{k}\bar{\mathbf{V}}_{f,m}\bar{\mathbf{V}}_{f,m}^{\dagger}\mathbf{H}_{k}^{\dagger}+\mathbf{H}_{k}\bar{\mathbf{\Omega}}\mathbf{H}_{k}^{\dagger}+\mathbf{\Sigma}_{\mathbf{z}_{k}}\right)\nonumber \\
 & -\log\det\left(\sum_{m=l+1}^{L}\mathbf{H}_{k}\bar{\mathbf{V}}_{f_{k},m}\bar{\mathbf{V}}_{f_{k},m}^{\dagger}\mathbf{H}_{k}^{\dagger}+\sum_{f\in\mathcal{F}_{\mathrm{req}}\setminus\{f_{k}\}}\sum_{m\in\mathcal{L}}\mathbf{H}_{k}\bar{\mathbf{V}}_{f,m}\bar{\mathbf{V}}_{f,m}^{\dagger}\mathbf{H}_{k}^{\dagger}+\mathbf{H}_{k}\bar{\mathbf{\Omega}}\mathbf{H}_{k}^{\dagger}+\mathbf{\Sigma}_{\mathbf{z}_{k}}\right),\nonumber
\end{align}
where we defined the notation $\bar{\mathbf{V}}\triangleq\{\bar{\mathbf{V}}_{f,l}\}_{f\in\mathcal{F}_{\mathrm{req}},\,l\in\mathcal{L}}$.

\subsection{Problem Definition and Optimization\label{sub:Problem-Formulation-Soft}}

As in Sec. \ref{sub:Problem-Formulation-Hard}, we aim at maximizing
the minimum-user rate $R_{\min}\triangleq\min_{f\in\mathcal{F}_{\mathrm{req}}}R_{f}$
subject to per-eRRH fronthaul capacity and transmit power constraints.
The problem is stated as\begin{subequations}\label{eq:problem-soft-original}
\begin{align}
\underset{\bar{\mathbf{V}},R_{\min},\mathbf{R}}{\mathrm{maximize}}\,\, & R_{\min}\label{eq:problem-soft-original-objective}\\
\mathrm{s.t.}\,\,\, & R_{\min}\leq\sum_{l\in\mathcal{L}}R_{f,l},\,\,f\in\mathcal{F}_{\mathrm{req}},\label{eq:problem-soft-original-rmin-constraint}\\
 & R_{f_{k},l}\leq q_{k,l}\left(\bar{\mathbf{V}},\mathbf{\Omega}\right),\,\,l\in\mathcal{L},\,k\in\mathcal{N}_{U},\label{eq:problem-soft-original-rate-constraint}\\
 & g_{i}\left(\bar{\mathbf{V}},\mathbf{\Omega}\right)\leq C_{i},\,\,i\in\mathcal{N}_{R},\label{eq:problem-soft-original-fronthaul-constraint}\\
 & R_{f,l}\leq S_{l},\,\,f\in\mathcal{F}_{\mathrm{req}},\,l\in\mathcal{L},\label{eq:problem-soft-original-file-size-constraint}\\
 & \sum_{f\in\mathcal{F}_{\mathrm{req}}}\sum_{l\in\mathcal{L}}\mathrm{tr}\left(\mathbf{E}_{i}^{\dagger}\bar{\mathbf{V}}_{f,l}\bar{\mathbf{V}}_{f,l}^{\dagger}\mathbf{E}_{i}+\mathbf{\Omega}_{i}\right)\leq P_{i},\,\,i\in\mathcal{N}_{R},\label{eq:problem-soft-original-power-constraint}
\end{align}
\end{subequations}where the function $g_{i}(\bar{\mathbf{V}},\mathbf{\Omega})$
is defined, with a small abuse of notation, from (\ref{eq:fronthaul-constraint-soft}),
as
\begin{align}
g_{i}\left(\bar{\mathbf{V}},\mathbf{\Omega}\right) & \triangleq\log\det\left(\sum_{f\in\mathcal{F}_{\mathrm{req}}}\sum_{l\in\mathcal{L}}\bar{c}_{f,l}^{i}\mathbf{E}_{i}^{\dagger}\bar{\mathbf{V}}_{f,l}\bar{\mathbf{V}}_{f,l}^{\dagger}\mathbf{E}_{i}+\mathbf{\Omega}_{i}\right)-\log\det\left(\mathbf{\Omega}_{i}\right),\label{eq:fronthaul-function-soft-modified}
\end{align}
given that, if $\bar{c}_{f,l}^{i}=1$, then $\mathbf{E}_{i}^{\dagger}\bar{\mathbf{V}}_{f,l}=\mathbf{U}_{f,l}^{i}$.

As for problem (\ref{eq:problem-hard-original}), we tackle (\ref{eq:problem-soft-original})
by means of the CCCP approach as applied to a rank-relaxed version
of (\ref{eq:problem-hard-original}), where the optimization variables
are given as $\mathbf{W}_{f,l}\triangleq\bar{\mathbf{V}}_{f,l}\bar{\mathbf{V}}_{f,l}^{\dagger}$
and the rank constraints $\mathrm{rank}(\mathbf{W}_{f,l})\leq n_{S,f,l}$
are relaxed. The resulting algorithm is detailed in Algorithm 2, where
we defined the functions
\begin{align}
 & \tilde{q}_{k,l}\left(\mathbf{W}^{(t+1)},\mathbf{\Omega}^{(t+1)},\mathbf{W}^{(t)},\mathbf{\Omega}^{(t)}\right)\label{eq:achievable-rate-linearized-soft}\\
\triangleq & \log\det\left(\sum_{m=l}^{L}\mathbf{H}_{k}\mathbf{W}_{f_{k},m}^{(t+1)}\mathbf{H}_{k}^{\dagger}+\sum_{f\in\mathcal{F}_{\mathrm{req}}\setminus\{f_{k}\}}\sum_{m\in\mathcal{L}}\mathbf{H}_{k}\mathbf{W}_{f,m}^{(t+1)}\mathbf{H}_{k}^{\dagger}+\mathbf{H}_{k}\bar{\mathbf{\Omega}}^{(t+1)}\mathbf{H}_{k}^{\dagger}+\mathbf{\Sigma}_{\mathbf{z}_{k}}\right)\nonumber \\
 & -\varphi\left(\begin{array}{c}
\sum_{m=l+1}^{L}\mathbf{H}_{k}\mathbf{W}_{f_{k},m}^{(t+1)}\mathbf{H}_{k}^{\dagger}+\sum_{f\in\mathcal{F}_{\mathrm{req}}\setminus\{f_{k}\}}\sum_{m\in\mathcal{L}}\mathbf{H}_{k}\mathbf{W}_{f,m}^{(t+1)}\mathbf{H}_{k}^{\dagger}\\
+\mathbf{H}_{k}\bar{\mathbf{\Omega}}^{(t+1)}\mathbf{H}_{k}^{\dagger}+\mathbf{\Sigma}_{\mathbf{z}_{k}},\\
\sum_{m=l+1}^{L}\mathbf{H}_{k}\mathbf{W}_{f_{k},m}^{(t)}\mathbf{H}_{k}^{\dagger}+\sum_{f\in\mathcal{F}_{\mathrm{req}}\setminus\{f_{k}\}}\sum_{m\in\mathcal{L}}\mathbf{H}_{k}\mathbf{W}_{f,m}^{(t)}\mathbf{H}_{k}^{\dagger}\\
+\mathbf{H}_{k}\bar{\mathbf{\Omega}}^{(t)}\mathbf{H}_{k}^{\dagger}+\mathbf{\Sigma}_{\mathbf{z}_{k}}
\end{array}\right),\nonumber \\
\mathrm{and}\,\, & \tilde{g}_{i}\left(\mathbf{W}^{(t+1)},\mathbf{\Omega}^{(t+1)},\mathbf{W}^{(t)},\mathbf{\Omega}^{(t)}\right)\label{eq:fronthaul-function-linearized-soft}\\
\triangleq & \varphi\left(\begin{array}{c}
\sum_{f\in\mathcal{F}_{\mathrm{req}}}\sum_{l\in\mathcal{L}}\bar{c}_{f,l}^{i}\mathbf{E}_{i}^{\dagger}\mathbf{W}_{f,l}^{(t+1)}\mathbf{E}_{i}+\mathbf{\Omega}_{i}^{(t+1)},\\
\sum_{f\in\mathcal{F}_{\mathrm{req}}}\sum_{l\in\mathcal{L}}\bar{c}_{f,l}^{i}\mathbf{E}_{i}^{\dagger}\mathbf{W}_{f,l}^{(t)}\mathbf{E}_{i}+\mathbf{\Omega}_{i}^{(t)}
\end{array}\right)-\log\det\left(\mathbf{\Omega}_{i}^{(t+1)}\right).\nonumber
\end{align}
After the convergence of the algorithm, each precoding matrix $\bar{\mathbf{V}}_{f,l}$
is obtained as $\bar{\mathbf{V}}_{f,l}\leftarrow\mathbf{V}_{n_{S,f,l}}(\mathbf{W}_{f,l})\mathrm{diag}(\mathbf{d}_{n_{S,f,l}}(\mathbf{W}_{f,l}))^{1/2}$
as in Sec. \ref{sub:Problem-Formulation-Hard}.

\begin{algorithm}
\caption{CCCP algorithm for problem (\ref{eq:problem-soft-original})}

\textbf{1.} Initialize the matrices $\mathbf{W}^{(1)}$ and $\mathbf{\Omega}^{(1)}$
to arbitrary positive semidefinite matrices that satisfy the per-eRRH
fronthaul capacity constraints (\ref{eq:problem-soft-original-fronthaul-constraint})
and power constraints (\ref{eq:problem-soft-original-power-constraint})
and set $t=1$.

\textbf{2.} Update the matrices $\mathbf{W}^{(t+1)}$ and $\mathbf{\Omega}^{(t+1)}$
as a solution of the following convex problem:\begin{subequations}\label{eq:problem-soft-DC}
\begin{align}
\underset{\mathbf{W}^{(t+1)},\mathbf{\Omega}^{(t+1)}\succeq\mathbf{0},R_{\min},\mathbf{R}}{\mathrm{maximize}}\,\, & R_{\min}\label{eq:problem-soft-DC-objective}\\
\mathrm{s.t.}\,\,\, & R_{\min}\leq\sum_{l\in\mathcal{L}}R_{f,l},\,\,f\in\mathcal{F}_{\mathrm{req}},\label{eq:problem-soft-DC-rmin-constraint}\\
 & R_{f_{k},l}\leq\tilde{q}_{k,l}\left(\mathbf{W}^{(t+1)},\mathbf{\Omega}^{(t+1)},\mathbf{W}^{(t)},\mathbf{\Omega}^{(t)}\right),\,\,l\in\mathcal{L},\,k\in\mathcal{N}_{U},\label{eq:problem-soft-DC-rate-constraint}\\
 & \tilde{g}_{i}\left(\mathbf{W}^{(t+1)},\mathbf{\Omega}^{(t+1)},\mathbf{W}^{(t)},\mathbf{\Omega}^{(t)}\right)\leq C_{i},\,\,i\in\mathcal{N}_{R},\label{eq:problem-soft-DC-fronthaul-constraint}\\
 & R_{f,l}\leq S_{l},\,\,f\in\mathcal{F}_{\mathrm{req}},\,l\in\mathcal{L},\label{eq:problem-soft-DC-file-size-constraint}\\
 & \sum_{f\in\mathcal{F}_{\mathrm{req}}}\sum_{l\in\mathcal{L}}\mathrm{tr}\left(\mathbf{E}_{i}^{\dagger}\mathbf{W}_{f,l}^{(t+1)}\mathbf{E}_{i}+\mathbf{\mathbf{\Omega}}_{i}^{(t+1)}\right)\leq P_{i},\,\,i\in\mathcal{N}_{R},\label{eq:problem-soft-DC-power-constraint}
\end{align}
\end{subequations}

\textbf{3.} Stop if a convergence criterion is satisfied. Otherwise,
set $t\leftarrow t+1$ and go back to Step 2.
\end{algorithm}

\section{Delivery Phase With Hybrid Fronthauling\label{sec:Hybrid-Fronthauling}}

In this section, we consider the design of a hybrid hard- and soft-transfer
mode fronthauling scheme, whereby, unlike the strategies discussed
in Sec. \ref{sec:Hard-Transfer} and Sec. \ref{sec:Soft-Transfer},
the capacity of each fronthaul link is generally used to carry both
hard and soft information about the uncached files. A similar scheme
was also considered in \cite{Patil-Yu} for a system with no caching.
In this scheme, as a hybrid of (\ref{eq:linear-precoding-hard}) or
(\ref{eq:transmitted-signal-soft-transfer}), the signal $\mathbf{x}_{i}$
transmitted by eRRH $i$ on the downlink channel is given as
\begin{equation}
\mathbf{x}_{i}=\sum_{f\in\mathcal{F}_{\mathrm{req}}}\sum_{l\in\mathcal{L}}\left(1-\bar{c}_{f,l}^{i}\bar{d}_{f,l}^{i}\right)\mathbf{V}_{f,l}^{i}\mathbf{s}_{f,l}+\hat{\mathbf{x}}_{i},\label{eq:transmitted-signal-hybrid-transfer}
\end{equation}
where, as for (\ref{eq:transmitted-signal-soft-transfer}), $\mathbf{V}_{f,l}^{i}\in\mathbb{C}^{n_{R,i}\times n_{S,f,l}}$
is the precoding matrix applied by eRRH $i$ on the baseband signal
$\mathbf{s}_{f,l}$ encoding the subfile $(f,l)$, and $\hat{\mathbf{x}}_{i}$
represents the quantized baseband signal received from the BBU on
the fronthaul link. Similar to (\ref{eq:transmitted-signal-soft-transfer}),
the first term for subfile $(f,l)$ is non-zero if the subfile $(f,l)$
is available at the eRRH by caching or via hard-mode fronthauling,
i.e., with $c_{f,l}^{i}=1$ or $d_{f,l}^{i}=1$, respectively.

The BBU precodes the subfiles $(f,l)$ that are not available at eRRH
$i$, i.e., with $\bar{c}_{f,l}^{i}\bar{d}_{f,l}^{i}=1$, producing
the signal
\begin{equation}
\tilde{\mathbf{x}}_{i}=\sum_{f\in\mathcal{F}_{\mathrm{req}}}\sum_{l\in\mathcal{L}}\bar{c}_{f,l}^{i}\bar{d}_{f,l}^{i}\mathbf{U}_{f,l}^{i}\mathbf{s}_{f,l},\label{eq:precoding-BBU-hybrid}
\end{equation}
where $\mathbf{U}_{f,l}^{i}\in\mathbb{C}^{n_{R,i}\times n_{S,f,l}}$
is the precoding matrix for the baseband signal $\mathbf{s}_{f,l}$.
The quantized signal $\hat{\mathbf{x}}_{i}$ in the right-hand side
of (\ref{eq:transmitted-signal-hybrid-transfer}) is given as (\ref{eq:quantized-signal-soft})
which can be reliably recovered by eRRH $i$ if the condition
\begin{align}
g_{i}\left(\mathbf{U},\mathbf{\Omega}\right) & \triangleq I\left(\tilde{\mathbf{x}}_{i};\hat{\mathbf{x}}_{i}\right)\label{eq:decompression-constraint-hybrid}\\
 & =\log\det\left(\sum_{f\in\mathcal{F}_{\mathrm{req}}}\sum_{l\in\mathcal{L}}\bar{c}_{f,l}^{i}\bar{d}_{f,l}^{i}\mathbf{U}_{f,l}^{i}\mathbf{U}_{f,l}^{i\dagger}+\mathbf{\Omega}_{i}\right)-\log\det\left(\mathbf{\Omega}_{i}\right)\leq\tilde{C}_{i}\nonumber
\end{align}
is satisfied, where we recall that $\mathbf{\Omega}_{i}$ denotes
the covariance matrix of the quantization noise in (\ref{eq:quantized-signal-soft}),
and we defined $\tilde{C}_{i}\leq C_{i}$ as the rate used on the
$i$th fronthaul for the soft-transfer mode. The rest of the frontahul
link of $C_{i}-\tilde{C}_{i}$ bit/symbol can be used for the hard-transfer
mode, i.e., for transferring the subfiles $(f,l)$ with $d_{f,l}^{i}=1$.
Accounting for both soft- and hard-transfer fronthauling, the fronthaul
capacity constraint for each eRRH $i$ is then stated as
\begin{equation}
\sum_{f\in\mathcal{F}}\sum_{l\in\mathcal{L}}d_{f,l}^{i}R_{f,l}+\tilde{C}_{i}\leq C_{i}.\label{eq:fronthaul-capacity-constraint-hybrid-transfer}
\end{equation}

With (\ref{eq:transmitted-signal-hybrid-transfer}), the signal $\mathbf{y}_{k}$
received by UE $k$ in (\ref{eq:received-signal}) can be written
as (\ref{eq:received-signal-soft}), with the only difference that
the aggregated precoding matrix $\bar{\mathbf{V}}_{f,l}\triangleq[\bar{\mathbf{V}}_{f,l}^{1};\ldots;\bar{\mathbf{V}}_{f,l}^{N_{R}}]$
for subfile $(f,l)$ consists of the submatrices $\bar{\mathbf{V}}_{f,l}^{i}\triangleq(1-\bar{c}_{f,l}^{i}\bar{d}_{f,l}^{i})\mathbf{V}_{f,l}^{i}+\bar{c}_{f,l}^{i}\bar{d}_{f,l}^{i}\mathbf{U}_{f,l}^{i}$.
Assuming the SIC decoding with the same decoding order, the rate $R_{f_{k},l}$
of the subfile $(f_{k},l)$ is achievable if the condition (\ref{eq:achievable-rate-soft})
is satisfied.

\subsection{Problem Definition and Optimization\label{sub:Problem-Definition-Hybrid}}

We aim at optimizing the precoding matrices $\mathbf{V}$ and $\mathbf{U}$
applied at the eRRHs and the BBU, along with the capacities $\tilde{\mathbf{C}}\triangleq\{\tilde{C}_{i}\}_{i\in\mathcal{N}_{R}}$
used for soft-transfer fronthauling, with the goal of maximizing the
minimum-user rate, as in Sec. \ref{sub:Problem-Formulation-Hard}
and Sec. \ref{sub:Problem-Formulation-Soft}, while satisfying the
fronthaul capacity (\ref{eq:fronthaul-capacity-constraint-hybrid-transfer})
and per-eRRH power constraints (\ref{eq:per-eRRH-power-constraint}).
The problem can be formulated as\begin{subequations}\label{eq:problem-hybrid-original}
\begin{align}
\underset{\bar{\mathbf{V}},R_{\min},\mathbf{R},\tilde{\mathbf{C}}}{\mathrm{maximize}}\,\, & R_{\min}\label{eq:problem-hybrid-original-objective}\\
\mathrm{s.t.}\,\,\, & R_{\min}\leq\sum_{l\in\mathcal{L}}R_{f,l},\,\,f\in\mathcal{F}_{\mathrm{req}},\label{eq:problem-hybrid-original-rmin-constraint}\\
 & R_{f_{k},l}\leq q_{k,l}\left(\bar{\mathbf{V}},\mathbf{\Omega}\right),\,\,l\in\mathcal{L},\,k\in\mathcal{N}_{U},\label{eq:problem-hybrid-original-rate-constraint}\\
 & g_{i}\left(\bar{\mathbf{V}},\mathbf{\Omega}\right)\leq\tilde{C}_{i},\,\,i\in\mathcal{N}_{R},\label{eq:problem-hybrid-original-decompression-constraint}\\
 & \sum_{f\in\mathcal{F}}\sum_{l\in\mathcal{L}}d_{f,l}^{i}R_{f,l}+\tilde{C}_{i}\leq C_{i},\,\,i\in\mathcal{N}_{R},\label{eq:problem-hybrid-original-fronthaul-constraint}\\
 & R_{f,l}\leq S_{l},\,\,f\in\mathcal{F}_{\mathrm{req}},\,l\in\mathcal{L},\label{eq:problem-hybrid-original-file-size-constraint}\\
 & \sum_{f\in\mathcal{F}_{\mathrm{req}}}\sum_{l\in\mathcal{L}}\mathrm{tr}\left(\mathbf{E}_{i}^{\dagger}\bar{\mathbf{V}}_{f,l}\bar{\mathbf{V}}_{f,l}^{\dagger}\mathbf{E}_{i}+\mathbf{\Omega}_{i}\right)\leq P_{i},\,\,i\in\mathcal{N}_{R}.\label{eq:problem-hybrid-original-power-constraint}
\end{align}
\end{subequations}As for problems (\ref{eq:problem-hard-original})
and (\ref{eq:problem-soft-original}), we can apply the CCCP approach
to a rank-relaxed version of the problem (\ref{eq:problem-hybrid-original}),
where the rank constraints $\mathrm{rank}(\mathbf{W}_{f,l})\leq n_{S,f,l}$
are removed. The procedure follows in the same manner as for Algorithms
1 and 2, and will not be detailed here.

\section{Numerical Results\label{sec:Numerical-Results}}

\begin{figure}
\centering\includegraphics[width=12cm,height=9cm]{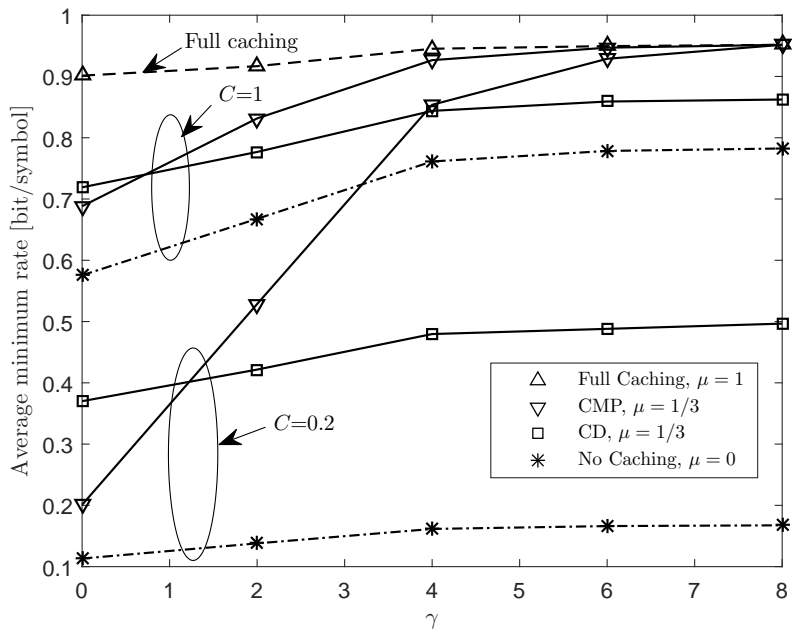}

\caption{\label{fig:graph-as-gamma}Average minimum rate $R_{\min}$ versus
the parameter $\gamma$ of the Zipf's distribution in (\ref{eq:Zipf-distribution})
for a F-RAN downlink under soft-transfer fronthauling mode ($\mu=0,1/3,1$,
$F=3$, $S=1$, $C=0.2$ and $1$ and $P/N_{0}=20$ dB).}
\end{figure}

In this section, we present some numerical results that compare the
performance of hard-transfer and soft-transfer fronthauling modes,
as well as of the hybrid scheme, with the pre-fetching strategies
discussed in Sec. \ref{sec:Pre-Fetching}. We consider an F-RAN system
where the positions of eRRHs and UEs are uniformly distributed within
a circular cell of radius $500$m. The channel $\mathbf{H}_{k,i}$
from eRRH $i$ to UE $k$ is modeled as $\mathbf{H}_{k,i}=\sqrt{\rho_{k,i}}\tilde{\mathbf{H}}_{k,i}$,
where the channel power $\rho_{k,i}$ is given as $\rho_{k,i}=1/(1+(d_{k,i}/d_{0})^{\alpha})$
and the elements of $\tilde{\mathbf{H}}_{k,i}$ are independent and
identically distributed (i.i.d.) as $\mathcal{CN}(0,1)$. We set the
parameters $d_{0}=50$m and $\alpha=3$. We consider a symmetric setting
where the covariance matrix $\mathbf{\Sigma}_{\mathbf{z}_{k}}$ is
given as $\mathbf{\Sigma}_{\mathbf{z}_{k}}=N_{0}\mathbf{I}$ for all
UEs $k\in\mathcal{N}_{U}$, and the eRRHs have the same transmit power
and fronthaul capacity, i.e., $P_{i}=P$ and $C_{i}=C$ for $i\in\mathcal{N}_{R}$
and are equipped with caches of equal size, i.e., $B_{i}=B$ and $\mu_{i}=\mu$
for $i\in\mathcal{N}_{R}$. For hard-transfer fronthauling, we assume
that the fronthaul transfer variables $\{d_{f,l}^{i}\}_{f\in\mathcal{F}_{\mathrm{req}},l\in\mathcal{L}}$
are set such that the subfile $(f_{k},l)$ requested by UE $k$ is
transferred on the fronthaul links to the $N_{F}$ eRRHs that have
the largest channel gains $||\mathbf{H}_{k,i}||_{F}^{2}$ to the UE
and have not stored the subfile, where $N_{F}\leq N_{R}$ is a parameter
that defines the scheme. Note that this implies that the cooperative
cluster of eRRHs for the transmission of any subfile for the hard-transfer
mode is of size $N_{F}$ plus the number of eRRHs that cache that
subfile. Moreover, the variables $\{d_{f,l}^{i}\}_{f\in\mathcal{F}_{\mathrm{req}},l\in\mathcal{L}}$
of the hybrid fronthauling strategy proposed in Sec. \ref{sec:Hybrid-Fronthauling}
is set to those of the hard-transfer mode with $N_{F}$ giving the
best performance. If not stated otherwise, we set $N_{R}=N_{U}=3$
and $n_{R,i}=n_{U,k}=1$.

We first study the impact of the file popularity on the F-RAN performance.
To this end, in Fig. \ref{fig:graph-as-gamma}, we plot the average
minimum rate $R_{\min}$ versus the parameter $\gamma$ of the Zipf's
distribution in (\ref{eq:Zipf-distribution}), where the average is
taken with respect to the channel, UEs' requests and the system geometry,
for an F-RAN downlink with soft-transfer fronthauling. We set the
parameters $F=3$, $S=1$, $C=0.2$ and $C=1$ and $P/N_{0}=20$ dB.
We compare the performance of CMP and CD pre-fetching with $\mu=1/3$
with the case of full ($\mu=1$) and no ($\mu=0$) caching (FCD is
not shown here to avoid clutter). Note that full caching is equivalent
to the MIMO broadcast part of the cut-set upper bound \cite[Theorem 14.10.1]{Cover}.
It is observed from the figure that the performance gain of the CMP
pre-fetching strategy with a larger $\gamma$, and hence with an increased
bias towards the most popular files, is more pronounced for lower
values of the fronthaul capacity $C$. This is because, in the regime
of small $C$, cooperative transmission by means of cloud processing,
as in C-RAN, cannot compensate for the lack of cooperation opportunities
on the cached files that affects the CD approach. In contrast, when
$\gamma$ is sufficiently small, the CD strategy outperforms CMP approach,
which suffers from a significant number of cache misses, particularly
for low values of $C$. We also note that, when $\gamma$ is sufficiently
large, the performance of CMP approaches that of full caching scheme
even with a small fronthaul capacity, due to the high probability
that cooperative transmission across all eRRHs is possible based only
on the cached contents.

\begin{figure}
\centering\includegraphics[width=12cm,height=9cm]{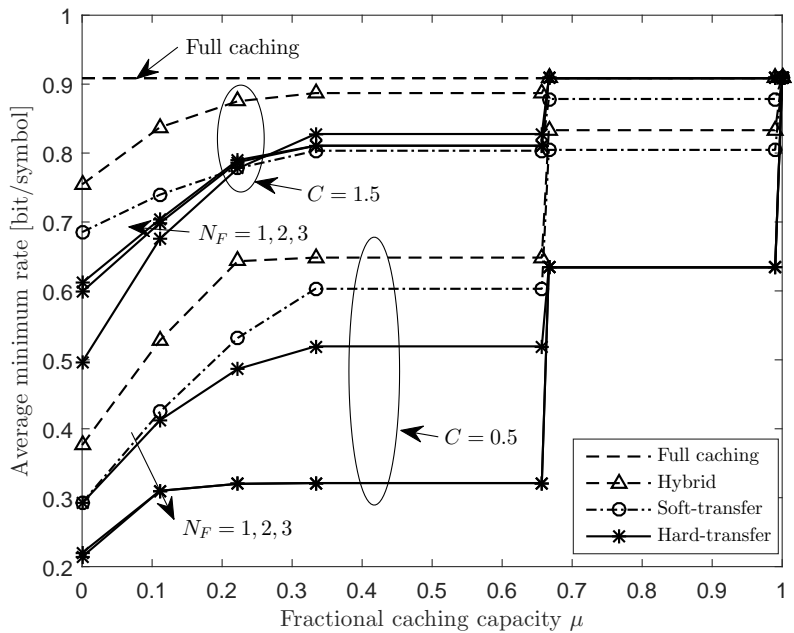}

\caption{\label{fig:graph-as-Mu}Average minimum rate $R_{\min}$ versus the
fractional caching capacity $\mu$ for an F-RAN downlink under FCD
pre-fetching ($C=0.5$ and $1.5$, $F=6$, $S=1$ and $P/N_{0}=20$
dB).}
\end{figure}

In Fig. \ref{fig:graph-as-Mu}, we investigate the effect of the fractional
caching capacity $\mu$ on the average minimum rate in two regimes
of fronthaul capacity, namely low, here, $C=0.5$ bit/symbol, and
moderate, here, $C=1.5$ bit/symbol. We adopt the FCD strategy and
compare the performance of soft- and hard-transfer fronthauling modes
with the hybrid mode proposed in Sec. \ref{sec:Hybrid-Fronthauling}.
Note that, as per the definition in Sec. \ref{sub:Fractional-Cache-Distinct},
FCD modifies its operation only at the values of $\mu=0$, $1/3$,
$2/3$ and $1$, which are marked in the figure. Note that all schemes
provide the same performance for $\mu=1$, since every eRRH has access
to the requested contents. The plot emphasizes the different relative
behavior of the soft and hard fronthauling strategies in different
fronthaul and caching set-ups. In particular, the soft-transfer fronthauling
strategy is seen to offer potentially large gains for low fronthaul
and sufficiently large caching capacities. This suggests that, if
the eRRHs have sufficient caching capabilities, soft-transfer fronthauling
provides the best way to use low-capacity fronthaul links. Conversely,
if the fronthaul capacity is large enough as compared to the minimum
delivery rate, and if the caching capacity is sufficiently large,
hard fronthauling can offer some, albeit not major, performance gains
over soft-mode fronthauling. We also observe that, for the hard-transfer
mode, the optimal size of the cooperative cluster, which depends on
$N_{F}$, increases with the fronthaul capacity. Finally, the hybrid
scheme is seen to outperform the soft- and hard-transfer modes, particularly
at lower caching capacities.

\begin{figure}
\centering\includegraphics[width=12cm,height=9cm]{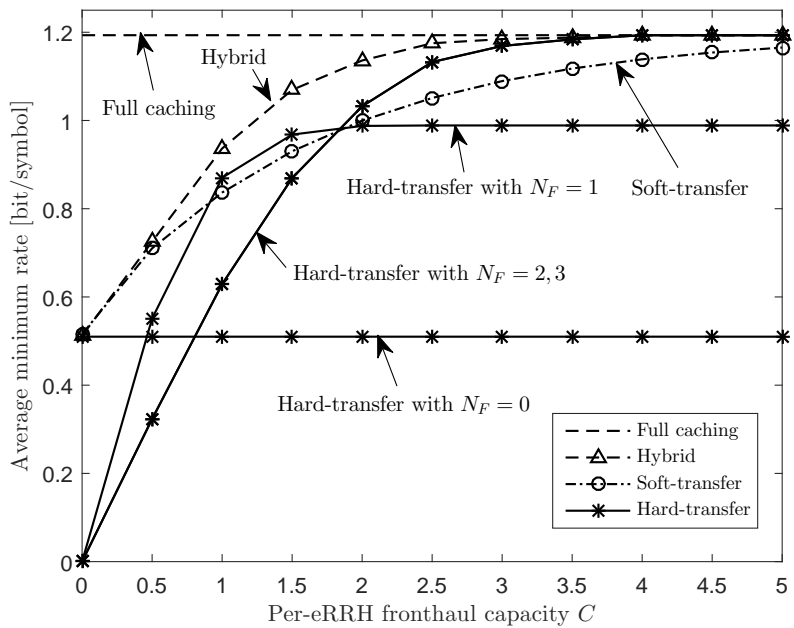}

\caption{\label{fig:graph-as-C}Average minimum rate $R_{\min}$ versus the
fronthaul capacity $C$ for an F-RAN downlink under FCD pre-fetching
($\mu=1/3$ and $1$, $F=6$, $S=2$, $\gamma=0.2$ and $P/N_{0}=20$
dB).}
\end{figure}

We then further study the role of the fronthaul capacity by plotting
in Fig. \ref{fig:graph-as-C} the average minimum rate $R_{\min}$
versus the fronthaul capacity $C$ for an F-RAN system with the FCD
pre-fetching, and with $\mu=1/3$ and $1$, $F=6$, $S=2$, $\gamma=0.2$
and $P/N_{0}=20$ dB. From the figure, we observe that the partial
caching capacity of the eRRHs, here with $\mu=1/3$, can be compensated
by a larger fronthaul capacity $C$. For instance, the soft-transfer
fronthauling mode with $\mu=1/3$ needs a fronthaul capacity of $C=3.38$
bit/symbol to achieve the full-caching upper bound within 5\%. Also,
it is seen that, for small fronthaul capacity $C$, it is desirable
to reduce the cluster size, and hence $N_{F}$, for hard-transfer
fronthauling, since a larger cluster size requires the transfer of
each subfile to more eRRHs on the fronthaul links of small capacity,
which limits the rate of the subfile. The figure confirms the observation
in Fig. \ref{fig:graph-as-Mu} that, if the fronthaul capacity $C$
is sufficiently large, the hard-transfer mode can provide some performance
gains over soft-transfer fronthauling, as long as the cooperative
cluster size is properly selected. Furthermore, we note that the hybrid
scheme has the capability to improve over both soft- and hard-mode
fronthauling, except for very low- and very high-fronthaul capacity
regime, in which it reverts to the soft- and hard-mode schemes, respectively.

\begin{figure}
\centering\includegraphics[width=12cm,height=9cm]{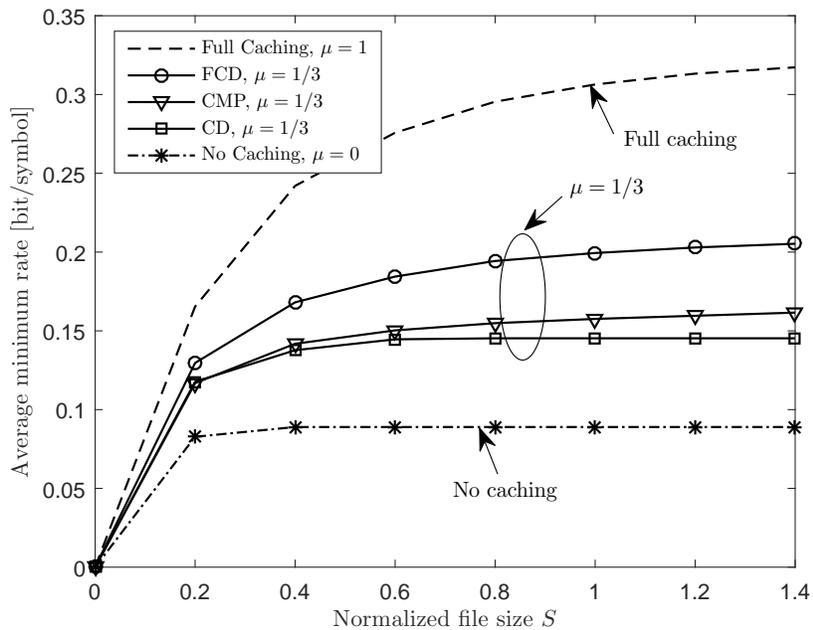}

\caption{\label{fig:graph-as-S}Average minimum rate $R_{\min}$ versus the
normalized file size $S$ for an F-RAN downlink under soft-transfer
mode fronthauling ($\mu=0,1/3$ and $1$, $F=6$, $C=0.5$, $\gamma=0.5$
and $P/N_{0}=10$ dB).}
\end{figure}

We now examine the impact of the file size $S$ on the optimal caching
policy. In Fig. \ref{fig:graph-as-S}, we show the average minimum
rate $R_{\min}$ versus the normalized file size $S$ for an F-RAN
downlink with soft-transfer mode fronthauling. We set the parameters
$F=6$, $C=0.5$, $\gamma=0.5$ and $P/N_{0}=10$ dB. The figure suggests
that, for all pre-fetching strategies, the minimum rate $R_{\min}$
increases with a larger $S$ in the regime of small file sizes, in
which the performance is limited by the file size $S$ rather than
the fronthaul capacity $C$. Moreover, the performance gain of the
FCD strategy compared to the CMP and CD is more pronounced for larger
$S$, since the partitioning of a file into multiple fragments becomes
more advantageous for the purpose of caching as the file size $S$
increases.

\begin{figure}
\centering\includegraphics[width=12cm,height=9cm]{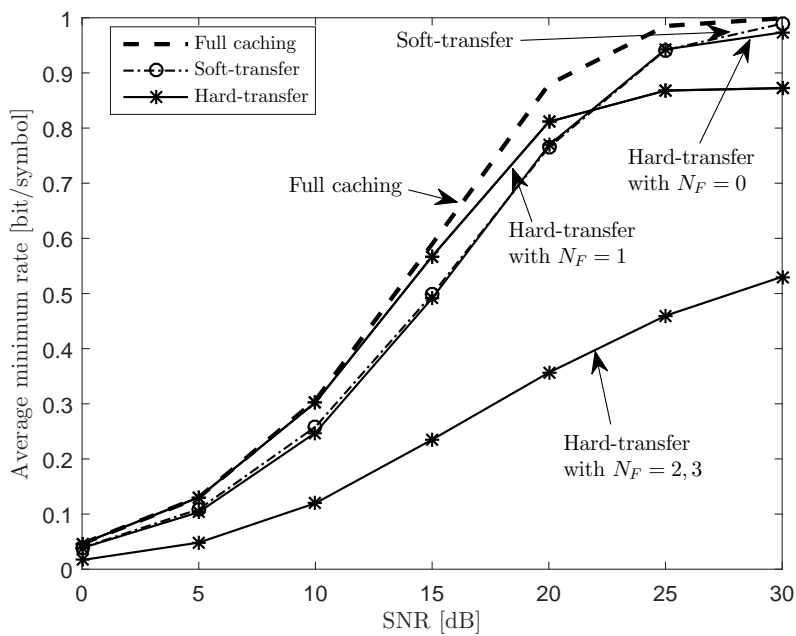}

\caption{\label{fig:graph-as-SNR}Average minimum rate $R_{\min}$ versus the
SNR $P/N_{0}$ for an F-RAN downlink under FCD pre-fetching ($\mu=1/3$
and $1$, $F=6$, $C=0.5$, $\gamma=0.5$ and $S=1$).}
\end{figure}

Finally, Fig. \ref{fig:graph-as-SNR} plots the average minimum rate
$R_{\min}$ versus the SNR $P/N_{0}$ for an F-RAN downlink with the
FCD pre-fetching and parameters set as $\mu=1/3$ and $1$, $F=6$,
$C=0.5$, $\gamma=0.5$ and $S=1$. It can be seen that, when the
SNR is large, the performance is limited by the fronthaul capacity
$C$, and thus increasing the cluster size of the hard-transfer fronthauling
results in a performance degradation. We can also see that soft-transfer
fronthauling, which has the flexibility to automatically control the
cluster size via the design of the precoding and quantization noises
covariance matrices, in this example, improves over the hard-transfer
scheme at sufficiently large SNRs.

\section{Conclusion\label{sec:Conclustion}}

In this work, we have studied joint design of cloud and edge processing
for an F-RAN architecture in which each edge node is equipped not
only with the functionalities of standard RRHs in C-RAN, but also
with local cache and baseband processing capabilities. For any given
pre-fetching strategy, we considered the optimization of the delivery
phase with the goal of maximizing the minimum delivery rate of the
requested files while satisfying the fronthaul capacity and per-eRRH
power constraints. We considered two basic fronthauling modes, namely
hard- and soft-transfer fronthauling, as well as a hybrid mode. Specifically,
with the hard-transfer mode, the fronthaul links are used to transmit
the requested files that are not in the local caches, while the soft-transfer
mode employs the fronthaul links following the C-RAN principle of
transferring quantized baseband signals. We compared the performance
of hard-, soft- and hybrid-transfer fronthauling modes with different
baseline pre-fetching strategies.

It was concluded, by means of extensive numerical results, that soft-transfer
provides a more effective way to use fronthaul resources than the
hard-transfer mode in most operating regimes except for very low SNR
regime and moderate fronthaul capacity. In such regimes, hard-transfer
fronthauling with a carefully selected cluster size can provide minor
gains. It is emphasized that these results hold under the assumptions
of information-theoretically optimal point-to-point compression for
communication on the fronthaul links. While it is known that point-to-point
compression can be improved upon \cite{Park-et-al:SPM}, the comparison
between the two modes should be revisited in the presence of less
effective compression or even only quantization (see also \cite{Kang-et-al}
for further discussion in the context of C-RAN). Moreover, the numerical
results highlighted the trade-off between fronthaul and caching resources,
whereby a smaller fronthaul capacity can be compensated for by a larger
cache, particularly for more skewed popularity distributions.

Among open problems, we mention here the analysis in the presence
of imperfect CSI and the design of a practical symbol-by-symbol, instead
of block, fronthaul quantization algorithms \cite{Lee-et-al}.

\end{document}